\documentclass[preprintnumbers,10pt,nofootinbib]{revtex4}

\usepackage{amsmath,latexsym,amssymb,amsfonts}
\usepackage[dvips]{graphicx,color}
\usepackage{bm}

\begin{document}


\title{\textbf{Entropy evolution of moving mirrors and the information loss problem}}

\author{
\textsc{Pisin Chen$^{a,b,c,d}$}\footnote{{\tt pisinchen{}@{}phys.ntu.edu.tw}} and \textsc{Dong-han Yeom$^{a}$}\footnote{{\tt innocent.yeom{}@{}gmail.com}}
}

\affiliation{
$^{a}$\small{Leung Center for Cosmology and Particle Astrophysics, National Taiwan University, Taipei 10617, Taiwan}\\
$^{b}$\small{Department of Physics, National Taiwan University, Taipei 10617, Taiwan}\\
$^{c}$\small{Graduate Institute of Astrophysics, National Taiwan University, Taipei 10617, Taiwan}\\
$^{d}$\small{Kavli Institute for Particle Astrophysics and Cosmology, SLAC National Accelerator Laboratory, Stanford University, Stanford, California 94305, USA}
}

\begin{abstract}
We investigate the entanglement entropy and the information flow of two-dimensional moving mirrors. Here we point out that various mirror trajectories can help to mimic different candidate resolutions to the information loss paradox following the semi-classical quantum field theory: (i) a suddenly stopping mirror corresponds to the assertion that all information is attached to the last burst, (ii) a slowly stopping mirror corresponds to the assertion that thermal Hawking radiation carries information, and (iii) a long propagating mirror corresponds to the remnant scenario. Based on such analogy, we find that the last burst of a black hole cannot contain enough information, while slowly emitting radiation can restore unitarity. For all cases, there is an apparent inconsistency between the picture based on quantum entanglements and that based on the semi-classical quantum field theory. Based on the quantum entanglement theory, a stopping mirror will generate a firewall-like violent emission which is in conflict with notions based on the semi-classical quantum field theory.
\end{abstract}

\maketitle

\newpage

\tableofcontents


\section{Introduction}

The black hole information loss problem \cite{Hawking:1976ra} is one of the most difficult yet essential paradoxes to be resolved toward the final theory of quantum gravity. The essence of the problem is this: how can we reconcile general relativity with unitary quantum mechanics. There is no commonly accepted answer yet, but here we provide a list of candidate resolutions to this problem \cite{Chen:2014jwq}.
\begin{itemize}
\item[1.] Information may not be conserved and the unitarity can be violated after evaporation \cite{Hawking:1976ra}. However, this is contradictory with current investigations on holography and AdS/CFT \cite{Maldacena:1997re}. Regarding this, there is still a debate, e.g., see \cite{Unruh:1995gn}.
\item[2.] Information can be preserved by Hawking radiation \cite{Susskind:1993if}. However, this may cause inconsistency \cite{Yeom:2008qw,Almheiri:2012rt}. Then, how can we overcome these paradoxes?
\item[3.] Information is not with the Hawking radiation but with other objects \cite{Chen:2014jwq}, e.g., via the following possibilities:
\begin{itemize}
\item[3-1.] A last burst of the evaporation carries information \cite{Hotta:2015yla} or the resolution/regularization of the singularity explains the whereabouts of the information \cite{Ashtekar:2005cj}.
\item[3-2.] A long-lifetime remnant carries all information \cite{Adler:2001vs}.
\item[3-3.] A bubble universe \cite{Farhi:1989yr} or large interior inside the black hole \cite{Christodoulou:2014yia} carries all information.
\item[3-4.] Some special quantum correlations can carry information \cite{Horowitz:2003he}.
\end{itemize}
Usual counter-arguments against 3-1 and 3-2 are related to the entropy bound \cite{Bousso:1999xy}. In general, the last burst, the last stage of the black hole evaporation, or the Planck scale objects have too small an amount of the Bekenstein-Hawking entropy. Usual counter-arguments against 3-3 and 3-4 are about their generality \cite{Chen:2014jwq}; these scenarios may not be applicable for the most general examples.
\item[4.] There may be other options, for examples, effective loss of information \cite{Sasaki:2014spa}, etc.
\end{itemize}

It is still unclear which is the correct answer. However, at least we can study their self-consistency by using some theoretical methods. One good toy model for this purpose is the moving mirror \cite{Davies:1976hi} (for historical remarks, see \cite{Elizalde}). The moving mirror is a surface that has the reflective boundary condition. Due to this boundary condition, as the mirror accelerates, it can create thermal particles that mimic Hawking radiation from a black hole \cite{Anderson:2015iga}. In addition, even though energy and entropy will be transmitted from the mirror, the causal structure is trivial and hence there is no way to lose unitarity. Therefore, we can see how the unitarity can be preserved by using this toy model and check which is the most probable idea among candidates of the resolution on the information loss problem \cite{Carlitz:1986nh,Wilczek:1993jn}.

One more fascinating point of moving mirrors is that they can be realized by tabletop experiments. Recently, Chen and Mourou \cite{Chen:2015bcg} suggested that accelerating plasma mirrors can be generated by plasma-laser interactions. If one impinges a strong laser pulse into a plasma medium, a plasma mirror with a controllable reflective index can be generated. By adjusting the densities of plasma layers, one can even control the speed or acceleration of the mirror. Then one can find a parameter space where the thermal radiation is detectable, through which various quantum field theoretical expectations can be validated or falsified. Of course, there remain experimental challenges, but regarding this potentially possible experiments, it is definitely worthwhile to provide theoretical guidances. In this paper, especially we will focus on the information loss problem. 

The most important step toward solving the information paradox is to understand the entanglement entropy. In the simplest case, the two-dimensional moving mirror, people already have investigated the entanglement entropy \cite{Holzhey:1994we,Fiola:1994ir}. Usually, the entanglement entropy depends on the cutoff scales. However, for two-dimensional systems with the conformal symmetry, one may apply the regularization and renormalization method to obtain a well-defined entanglement entropy, which was first shown by Holzhey, Larsen and Wilczek \cite{Holzhey:1994we} and was later simplified by Bianchi and Smerlak \cite{Bianchi:2014qua}. Thanks to this formula, we can see quantitative details on the restoration of unitarity by the radiation from the mirror. By adjusting different mirror trajectories, one can mimic different black hole evaporation scenarios associated with different resolutions \cite{Good:2013lca}, through which the self-consistency of these proposals can be examined.

This paper is organized as follows. In SEC.~\ref{sec:two}, we review the physics of two-dimensional moving mirrors and the entanglement entropy formula. In SEC.~\ref{sec:ent}, we show that the mirror can mimic different resolutions by adjusting the trajectory and it can be shown that some candidates are not suitable while some others are viable to explain the information loss problem. In SEC.~\ref{sec:amps}, we discuss more about the consistency and figure out the tension between the quantum entanglement and the semi-classical quantum field theory. Finally, in SEC.~\ref{sec:dis}, we summarize and discuss what one can learn and expect from the moving mirror experiments.

\section{\label{sec:two}Two dimensional moving mirrors}

Let us consider a two-dimensional moving mirror in a conformal field theory with the metric
\begin{eqnarray}
ds^{2} = - \alpha^{2}(u,v) dudv,
\end{eqnarray}
where $u$ is the retarded time (left-moving) and $v$ is the advanced time (right-moving). There is a mirror at $v = p(u)$, where we impose the reflective boundary condition on the mirror following the trajectory.

\begin{figure}
\begin{center}
\includegraphics[scale=0.65]{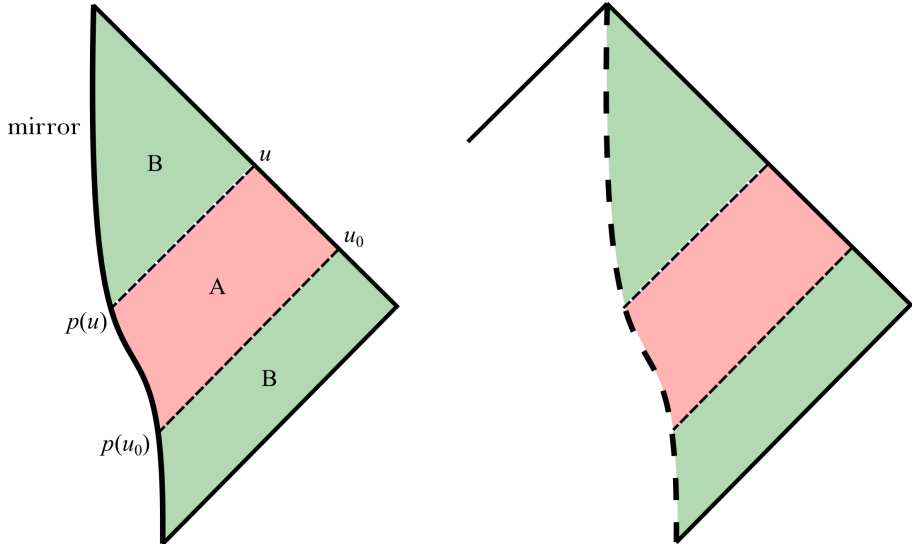}
\caption{\label{fig:mirror_concept}Left: the causal structure of a moving mirror. Right: if the mirror is partially reflective, then some information can be transmitted to the left boundary.}
\end{center}
\end{figure}

First we separate this space-time into two subsystems (left of FIG.~\ref{fig:mirror_concept}), where one corresponds to $A = [u_{0}, u]$, while the other, $B$, is its complementary. In two dimensions, every right-moving modes will be bounced by the mirror as one traces backward in time. Therefore, one can interpret that $A$ corresponds to the Hawking-like thermal radiation that is \textit{already} radiated from the mirror, while $B$ corresponds to the vacuum fluctuating modes complementary to the thermal radiation in $A$ that comove with the mirror, which are still \textit{not} radiated \textit{yet}.

Under this setting, it has been known that the entanglement entropy between $A$ and $B$ is formally defined from the density matrix $\rho_{A}$ for $A$, where
\begin{eqnarray}
\rho_{A} = \mathrm{Tr}_{B} \rho
\end{eqnarray}
and $\rho = | \Psi \rangle \langle \Psi |$ with $| \Psi \rangle$ the quantum state of the total system. Using this, the entanglement entropy is defined by
\begin{eqnarray}
S(A|B) = - \mathrm{Tr}_{A} \rho_{A} \log \rho_{A}.
\end{eqnarray}
For the two-dimensional conformal field theory, following Holzhey, Larsen and Wilczek \cite{Holzhey:1994we}, this can be evaluated by
\begin{eqnarray}
S(A|B) &=& \frac{c}{12} \log \frac{(u-u_{0})^{2}}{\epsilon^{2}}\\
&=& \frac{c}{12} \log \frac{(p(u)-p(u_{0}))^{2}}{p'(u) p'(u_{0}) \delta u \delta u_{0}},
\end{eqnarray}
where $c$ is the central charge that, for simplicity, can be taken as unity and $\epsilon$ is the UV-cutoff. Here, $\delta u$ and $\delta u_{0}$ both are introduced for the covariant regularization at $u$ and $u_{0}$ \cite{Bianchi:2014qua}, while both should approach zero and hence still there remains a divergence.

There is a cutoff dependence in this formula, but this can be avoided by subtracting a counter term. In Holzhey, Larsen and Wilczek \cite{Holzhey:1994we}, this term is subtracted by the corresponding term in the stationary mirror limit, i.e., $p(u) = u$ and $p(u_{0}) = u_{0}$, and they obtained the renormalized entropy
\begin{eqnarray}
S(A|B) = \frac{c}{12} \log \frac{(p(u)-p(u_{0}))^{2}}{(u-u_{0})^{2} p'(u) p'(u_{0})}.
\end{eqnarray}
Bianchi and Smerlak \cite{Bianchi:2014qua} further simplify this expression by choosing $u_{0} \rightarrow - \infty$ and they obtained
\begin{eqnarray}\label{eq:BS}
S(A|B) = - \frac{c}{12} \log p'(u).
\end{eqnarray}
The relation between the entanglement entropy $S$ and the position of the mirror as a function of $t$, say $x(t)$, can be presented by \cite{Good:2015nja}
\begin{eqnarray}
S(t) = -\frac{1}{6} \tanh^{-1} \dot{x}(t),
\end{eqnarray}
where $p'(u) = (1 + \dot{x}(t))/(1 - \dot{x}(t))$. By using this relation, one can further estimate the out-going energy flux according to the Davies-Fulling-Unruh formula \cite{Davies:1976ei} as a functional of the entanglement entropy \cite{Bianchi:2014qua,Good:2015nja}:
\begin{eqnarray}
F(t) = \frac{1}{2\pi} e^{-12 S} \left( 6 \dot{S}^{2} \tanh 6S + \ddot{S} \right) \cosh^{2} 6S.
\end{eqnarray}

There have been some attempts to generalize this result. First, one can generalize the entanglement entropy-flux relation for higher dimensional black hole systems \cite{Abdolrahimi:2015tha}. Eq.~(\ref{eq:BS}) works consistently except near the Page time and the end point of evaporation. Second, one can apply it to dynamical causal structures of black hole physics \cite{Bianchi:2014bma}. In this paper, we will not generalize this formula to various black holes (e.g., \cite{Abdolrahimi:2015tha,Bianchi:2014bma}) but focus on the moving mirror itself.

\section{\label{sec:ent}Entropy evolution by moving mirrors}

In this section, now we consider the entropy evolution associated with moving mirrors in two-dimensions. In addition to this, we will connect various trajectories of moving mirrors with candidate resolutions to the information loss problem.

Before we discuss details, in order to describe a closed system with finite number of states \cite{Page:1993wv}, we define and assume as follows:
\begin{itemize}
\item[--] Boltzmann entropy of the mirror side: $\bar{S}(M)$.
\item[--] Boltzmann entropy of the radiation: $\bar{S}(R)$.
\item[--] The total Boltzmann entropy of the radiation at $u = \infty$ is $\bar{S}_{\mathrm{m}}$, where we assume that $\bar{S}_{\mathrm{m}} = \log N$ is a constant for a closed system with a given constant $N$.
\end{itemize}

\begin{figure}
\begin{center}
\includegraphics[scale=0.65]{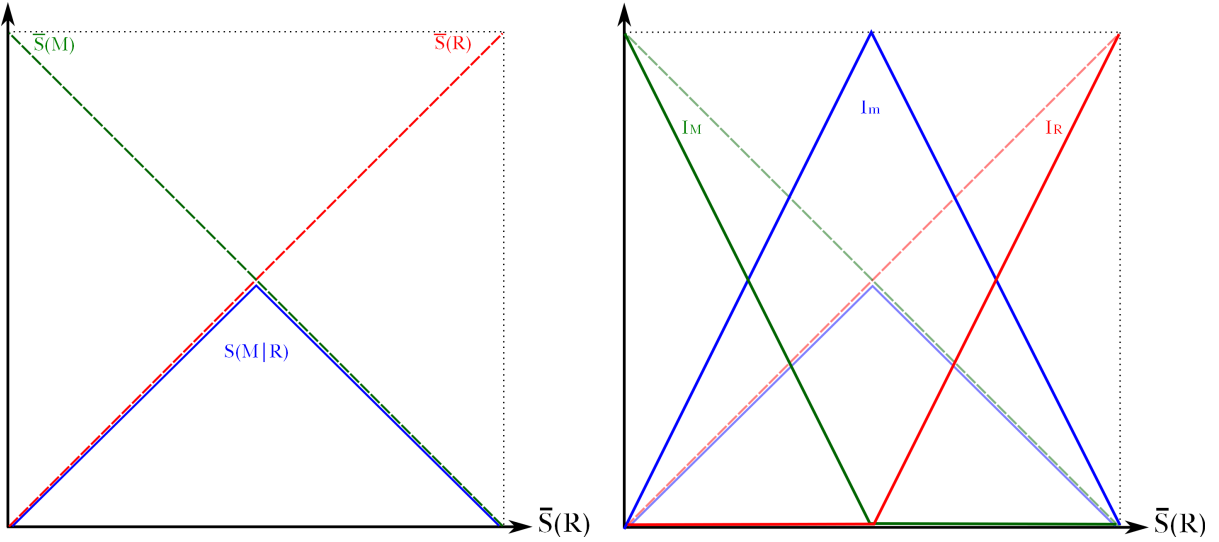}
\caption{\label{fig:PD}Entropy-information flow for a mirror.}
\end{center}
\end{figure}

Now we apply the entanglement entropy formula, where previous $A$ corresponds to $R$ and $B$ corresponds to $M$ in our new notations. The entanglement entropy and information will have the following properties:
\begin{itemize}
\item[--] The entanglement entropy is $S(R|M)$, where due to unitarity $S(R|M)=S(M|R)$.
\item[--] We can define the information measure of $M$ by $I_{M} = \bar{S}(M) - S(M|R)$.
\item[--] Radiation will generate the mutual information between mirror and future infinity, where it can be measured by $I_{\mathrm{m}} = S(R|M) + S(M|R) - S(M \cup R)$. Here, the entanglement entropy of the total system is $S(M \cup R) = 0$. Hence, for the unitary system, $I_{\mathrm{m}} = 2 S(R|M)$.
\item[--] We can define information of radiation $R$ at future infinity by $I_{R} = \bar{S}(R) - S(R|M)$.
\end{itemize}
Note that the total information is always conserved: $I = I_{M} + I_{\mathrm{m}} + I_{R} = \log N =$ const (FIG.~\ref{fig:PD}) \cite{Alonso-Serrano:2015bcr}. In terms of the entanglement entropy, after the mirror stops to move (i.e., $\dot{x} = 0$), it becomes zero again and hence we can regard that information is transmitted from the mirror $M$ to the future infinity by radiation $R$.

In two-dimensional cases, the black hole entropy is proportional to the mass, but the exact proportionality relation depends on the model parameters of the black hole system \cite{Callan:1992rs}. Therefore, the exact estimation of the thermal entropy via radiation (in terms of energy) is less clear. Keeping this in mind, we illustrate possible mirror histories as well as corresponding information transmission.

\subsection{Partially reflective mirror}

If the role of the mirror (the reflective boundary condition) fails by any reason, then we may not be able to apply this entanglement entropy formula. This possibility was discussed in \cite{Wilczek:1993jn} when the mirror is not totally reflective. This may mimic the bubble universe scenario (3-3) \cite{Farhi:1989yr} in the sense that the mirror takes a part of information (e.g., the left side of the future infinity, right of FIG.~\ref{fig:mirror_concept}) that cannot be accessed by the other side (e.g., the right side of the future infinity).

\subsection{Totally reflective mirror}

We use the following model:
\begin{eqnarray}
\frac{dS(t)}{dt} &=& A \sin^{2} \pi\frac{t}{t_{\mathrm{P}}} \;\;\;\;\;\;\;\;\;\;\;\;\;\;\;\;\;\;\;\;\;\;\;\;\;\;\;\;\;\; 0 \leq t < t_{\mathrm{P}},\\
&=& - A \frac{t_{\mathrm{P}}}{t_{\mathrm{f}}-t_{\mathrm{P}}} \sin^{2} \pi\frac{t-t_{\mathrm{P}}}{t_{\mathrm{f}}-t_{\mathrm{P}}} \;\;\;\;\;\;\;\; t_{\mathrm{P}} \leq t < t_{\mathrm{f}},
\end{eqnarray}
so that $S''(t)$ is continuous at $t_{\mathrm{P}}$ (may not be differentiable but this does not matter for later discussions). We choose $A = 1$ and $t_{\mathrm{P}} = 10$, remaining $t_{\mathrm{f}}$ as a free parameter. Of course, at once we fix a shape of $S(t)$, then we can reconstruct the corresponding trajectory $x(t)$ or $p(u)$.

Using this simplified model, we can name several different mirror trajectories, so-called (left of FIG.~\ref{fig:tot})
\begin{itemize}
\item[--] Suddenly stopping mirror: $t_{\mathrm{f}} = 15$,
\item[--] Slowly stopping mirror: $t_{\mathrm{f}} = 20$,
\item[--] Long propagating mirror: $t_{\mathrm{f}} = 50$.
\end{itemize}

\subsubsection{Suddenly stopping mirror}

For the suddenly stopping mirror case, it should release a very strong energy burst. As the decelerating time becomes shorter and shorter, the last burst energy increases more and more. Hence, it is not unreasonable to assume that the last burst can have enough capacity to restore all information within a short time.

\begin{figure}
\begin{center}
\includegraphics[scale=0.9]{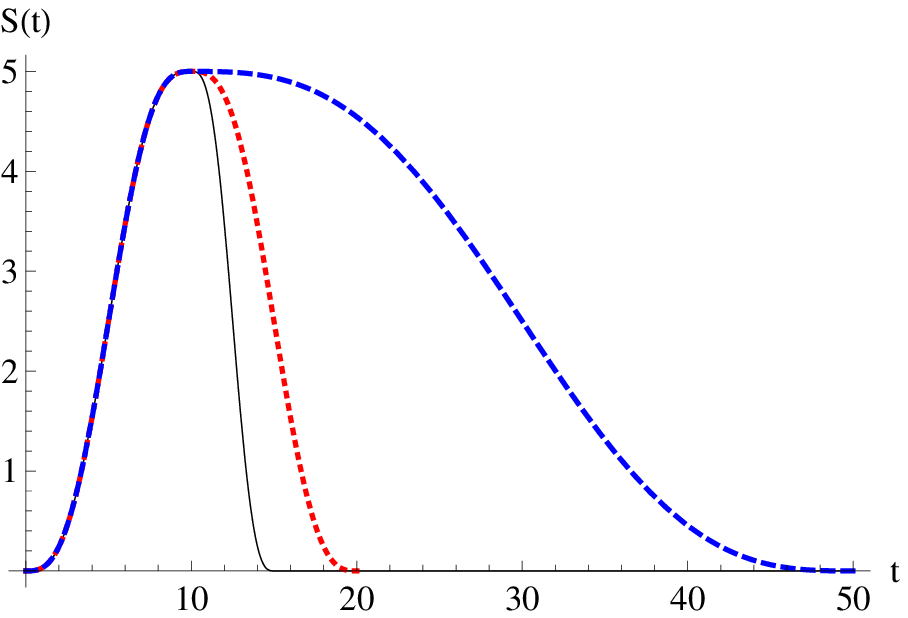}
\includegraphics[scale=0.9]{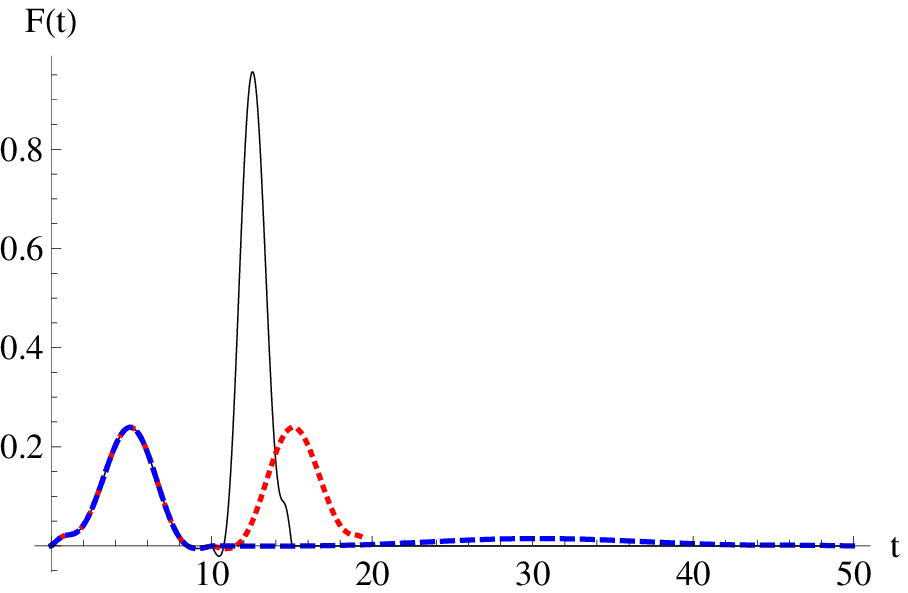}
\caption{\label{fig:tot} $S(t)$ and $F(t)$ for three cases (black: $t_{\mathrm{f}}=15$, red dotted: $t_{\mathrm{f}}=20$, blue dashed: $t_{\mathrm{f}}=50$).}
\end{center}
\end{figure}

However, it seems that the last burst of the moving mirror would emit too much energy (right of FIG.~\ref{fig:tot}). In the moving mirror case, this is not a problem, since the mirror is accelerating and this implies that there is an external source to provide the energy. The situation is different, however, in the case of the black hole. Due to the evaporation process, the last burst energy cannot exceed more than the Planck mass.

Therefore, this mirror trajectory \textit{cannot} realize the idea of 3-1 for black holes. If the final burst or emission happens during a very short time with a bounded energy \cite{Hotta:2015yla}, then it will not be helpful to explain the information loss problem.

\subsubsection{Slowly stopping mirror}

For the slowly stopping mirror case, one can imagine that after the Page time \cite{Page:1993wv}, the mirror begins to decrease its velocity. Then the radiation around the stopping phase transmits information, and hence the thermal or Hawking-like radiation will carry information. Typically, the radiated energy before the Page time and after the Page time should be of the similar order and hence this is consistent with the usual picture of black hole complementarity.

Then, the interesting question is whether there is any inconsistency, such as black hole complementarity \cite{Susskind:1993mu,Yeom:2008qw}. In the moving mirror case, there is no formal interior of the black hole and hence it may not be easy to do the same duplication experiment \cite{Susskind:1993mu}, but it will be still possible to apply the Almheiri-Marolf-Polchinski-Sully (AMPS) thought experiment \cite{Almheiri:2012rt}. We leave this topic to the next section.

\begin{figure}
\begin{center}
\includegraphics[scale=1]{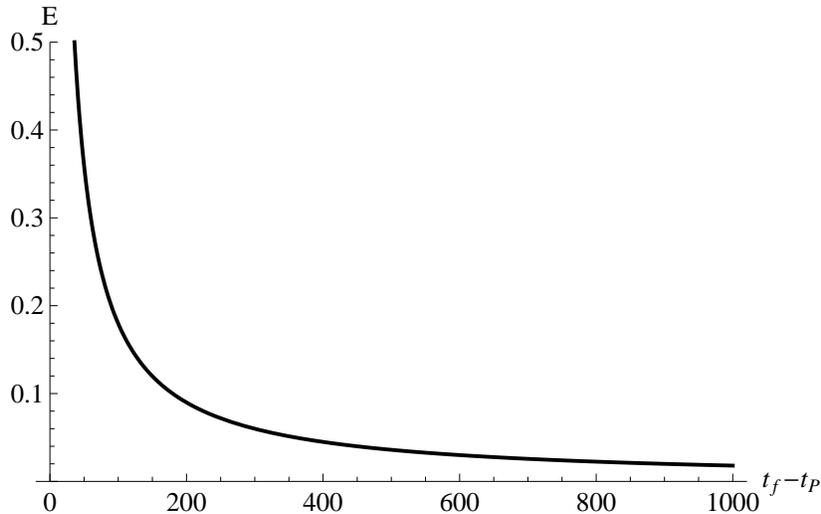}
\caption{\label{fig:energy} Total amount of out-going energy after the Page time: $E = \int_{t_{\mathrm{P}}}^{t_{\mathrm{f}}} F(t) (1-\dot{x}) dt$ as a function of $t_{\mathrm{f}} - t_{\mathrm{P}}$ (fixing $t_{\mathrm{P}} = 10$).}
\end{center}
\end{figure}

\subsubsection{Long propagating mirror}

For the long propagating mirror case, after the Page time, it takes a very long time to finish the deceleration \cite{Good:2015nja}. This mimics a very long lifetime remnant. One interesting observation is to see the total amount energy as a function of $t_{\mathrm{f}}$ (FIG.~\ref{fig:energy}). As the lifetime increases, we can reduce the radiated energy after the Page time. This means that if the late time radiation is slow enough, even though the radiated energy after the Page time is very small, it can contain enough information. So, this mimics the possibility that a negligible energy (e.g., vacuum) carries all correlations \cite{Hotta:2015yla} or a remnant preserves information \cite{Chen:2014jwq}.

There are some valuable comments. In the literature, there exist several models for two-dimensional dynamical black holes, especially by introducing a dilaton field. One famous example is the Callan-Giddings-Harvey-Strominger (CGHS) model \cite{Callan:1992rs}. In this model, we can introduce the exact form (up to the one-loop order) of the renormalized energy-momentum tensor \cite{Davies:1976ei} and can include semi-classical effects. By solving this system using numerical simulations \cite{Piran:1993tq,Ashtekar:2010hx}, we can see the detailed causal structures. One interesting numerical observation was that as the black hole finishes its evaporation, the total mass of the black hole approaches a universal constant \cite{Ashtekar:2010hx}. Perhaps, one can interpret that this limit implies the breakdown of the semi-classical approach. However, several authors suggested interpreting that this final universal constant mass implies a kind of Planck scale remnants \cite{Almheiri:2013wka}. Still, it is not possible to conclude, but at least our mirror calculation supports that such a remnant picture can functionally work in terms of the entanglement entropy.

There are some cautious comments for the generalization of this argument. First, we ignored non-perturbative effects. For the mirror case, it is less clear what is the role of non-perturbative effects. For black hole cases, the existence of remnants may cause the infinite production problem \cite{Chen:2014jwq} (although for two-dimensional cases, there is a support in \cite{Almheiri:2013wka}). Second, we considered only for two-dimensional mirrors. Perhaps, for black hole cases, as the dimension increases, such a universal end-point mass may not exist.

In summary, we \textit{cannot} conclude that the remnant picture is completely viable. However, we are certain that, at least for two-dimensions, the remnant picture is viable and it may even be the real answer to the information loss problem.

\section{\label{sec:amps}Consistency check: is there a firewall from a mirror?}

Up to this section, we have investigated the entanglement entropy and the energy flux released from the mirror. The former is related to the unitary quantum mechanics whereas the latter is related to the semi-classical quantum field theory. In previous sections, we implicitly assumed that these two aspects are consistent. But are they really consistent?

In this section, we first summarize the original version of the AMPS though experiment. We then apply similar thought experiments in the context of the mirror dynamics. We will conclude that such a thought experiment is indeed possible and this seems to reveal the tension between the quantum entanglement and the semi-classical quantum field theory even with a moving mirror.

\subsection{AMPS thought experiment: the original version}

Black hole complementarity assumes \cite{Susskind:1993if}:
\begin{description}
\item[--] A1. Unitarity. There exists an observer who can recover all information/correlation of collapsed matter into the event horizon.
\item[--] A2. For an asymptotic observer, the local quantum field theory, i.e., semi-classical descriptions on Hawking radiation is a good description.
\item[--] A3. For a free-falling observer, general relativity is a good description. Hence, the in-falling observer can probe inside the black hole until the observer touches the singularity.
\end{description}
In addition, although usually people do not explicitly present, we need further two assumptions \cite{Chen:2014jwq}:
\begin{description}
\item[--] A4. The area of the black hole is proportional to the coarse-grained entropy of the black hole. Hence, information begins to come out from the black hole by Hawking radiation around the Page time \cite{Page:1993wv} (still the black hole is semi-classical).
\item[--] A5. There is an ideal in-falling observer who satisfies A1-A4 and counts information/states.
\end{description}
Now black hole complementarity argues that these assumptions are consistent and no observer can see the violation of the previous assumptions.

\begin{figure}
\begin{center}
\includegraphics[scale=0.65]{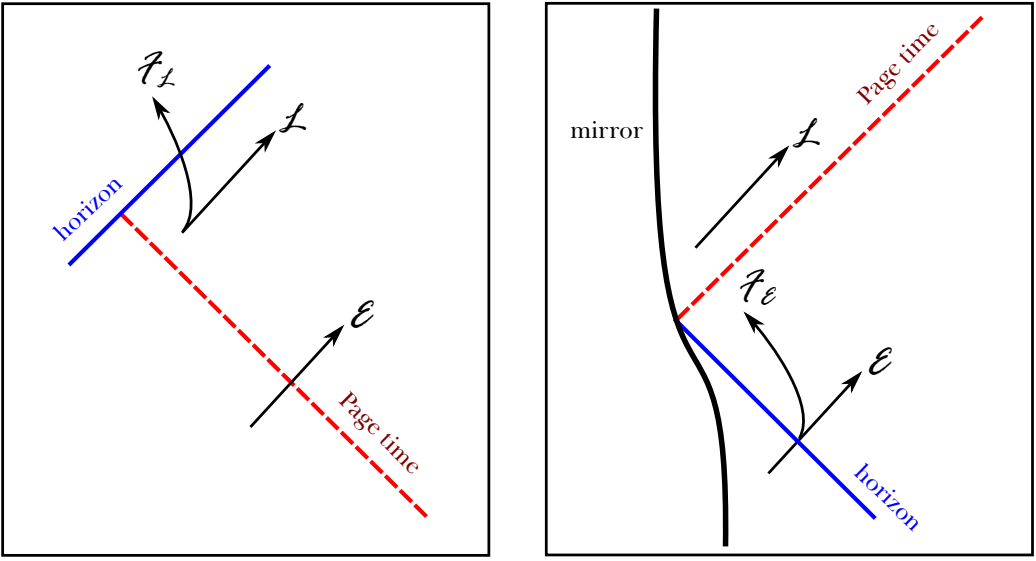}
\caption{\label{fig:amps}Left: the AMPS thought experiment for the original black hole version. Right: the AMPS thought experiment for the moving mirror.}
\end{center}
\end{figure}

Let us assume that a black hole is unitary and (after the Page time) Hawking radiation should contain information. If we denote that the earlier part (before the Page time) of radiation is $\mathcal{E}$ and the later part (after the Page time) of radiation is $\mathcal{L}$ (left of FIG.~\ref{fig:amps}), then the entanglement entropy between $\mathcal{E}$ and $\mathcal{L}$ should strictly decrease as the black hole evaporates. On the other hand, the internal degrees of freedom is also independent according to general relativity, unless there is a special event inside the horizon. Therefore, the in-falling counterpart of later Hawking radiation $\mathcal{F}_{\mathcal{L}}$ will form another independent system.

Then AMPS suggested to check the strong subadditivity \cite{Almheiri:2012rt}, where it should be satisfied in general:
\begin{eqnarray}
S_{\mathcal{E}\mathcal{L}} + S_{\mathcal{L}\mathcal{F}_{\mathcal{L}}} \geq S_{\mathcal{L}} + S_{\mathcal{E}\mathcal{L}\mathcal{F}_{\mathcal{L}}},
\end{eqnarray}
where $S_{\mathcal{X}}$ is the von Neumann entropy of the system $\mathcal{X}$. For two systems $\mathcal{X}$ and $\mathcal{Y}$, $S_{\mathcal{X}\mathcal{Y}} \equiv S_{\mathcal{X} \cup \mathcal{Y}}$. 
Keeping this in mind, by relying on A5, let us assume that there exists a free-falling observer $\mathcal{O}$ who satisfies A1 to A4. Then according to $\mathcal{O}$, the following logic should be true.
\begin{itemize}
\item[-- L1.] We know that after the Page time (by A1, A2, and A4), as the black hole evaporates, the entropy of the outside the black hole ($\mathcal{E} \cup \mathcal{L}$) should gradually decrease from the maximum value $S_{\mathcal{E}}$ to zero. 
Hence, $\mathcal{O}$ obtains $S_{\mathcal{E}\mathcal{L}} < S_{\mathcal{E}}$.
\item[-- L2.] If there is no special event for $\mathcal{O}$ (by A3), then a creation of a Hawking particle pair can be regarded as a localized unitary quantum evolution. Hence, $\mathcal{L} \cup \mathcal{F}_{\mathcal{L}}$ should be in a pure state and $S_{\mathcal{L}\mathcal{F}_{\mathcal{L}}} = 0$ for $\mathcal{O}$. From the subadditivity relation
\begin{eqnarray}
|S_{\mathcal{E}} - S_{\mathcal{L}\mathcal{F}_{\mathcal{L}}}| \leq S_{\mathcal{E}\mathcal{L}\mathcal{F}_{\mathcal{L}}} \leq S_{\mathcal{E}} + S_{\mathcal{L}\mathcal{F}_{\mathcal{L}}},
\end{eqnarray}
we obtain that $S_{\mathcal{E}\mathcal{L}\mathcal{F}_{\mathcal{L}}} = S_{\mathcal{E}}$.
\item[-- L3.] By plugging L1 and L2 to the strong subadditivity relation, we obtain the relation
\begin{eqnarray}
S_{\mathcal{E}} > S_{\mathcal{E}} + S_{\mathcal{L}},
\end{eqnarray}
where this is impossible.
\end{itemize}
Therefore, the assumptions of black hole complementarity are inconsistent, if there is an observer who counts $\mathcal{E}$ and $\mathcal{L}$ outside the black hole, falls into the black hole after the Page time, and check the strong subadditivity inside the horizon.

\subsection{AMPS thought experiment: the mirror version}

For the moving mirror case, A1, A2, A3, and A5 would be satisfied. It is less clear whether A4 is true or not, but thanks to the entanglement entropy formula, we can define the Page time. Then the question is this: can the moving mirror overcome inconsistency? Regarding this, the crucial point is as follows: before the Page time, the mirror should \textit{accelerate}, while after the Page time, the mirror should \textit{decelerate}.

Before the Page time, the mirror accelerates and there appears a \textit{horizon}. Hence, a separation between the in-going mode and the out-going mode are possible \cite{Hotta:2015yla} (right of FIG.~\ref{fig:amps}). Let us distinguish that the out-going radiation before the Page time is $\mathcal{E}$, while its in-going counterpart is $\mathcal{F}_\mathcal{E}$ \cite{Hotta:2015yla}. On the other hand, during the decelerating phase (after the Page time), there is no well-defined horizon and hence the separation between $\mathcal{L}$ and its counterpart $\mathcal{F}_{\mathcal{L}}$ are not well-defined.

Then, we can find three independent systems $\mathcal{E}$, $\mathcal{F}_{\mathcal{E}}$, and $\mathcal{L}$. Now we can apply the same analysis, so to speak L1$'$, L2$'$, and L3$'$ as follows.
\begin{itemize}
\item[-- L1$'$.] We know that after the Page time (by A1, A2, and A4), the entropy of the outside the mirror ($\mathcal{E} \cup \mathcal{L}$) should gradually decrease from the maximum value $S_{\mathcal{E}}$ to zero. Hence, $S_{\mathcal{L}\mathcal{E}} < S_{\mathcal{E}}$.
\item[-- L2$'$.] If there is no special event for $\mathcal{O}$ (by A3), then a creation of a Hawking particle pair can be regarded as a localized unitary quantum evolution. Hence, $\mathcal{E} \cup \mathcal{F}_{\mathcal{E}}$ should be in a pure state and $S_{\mathcal{E}\mathcal{F}_{\mathcal{E}}} = 0$. From the subadditivity relation
\begin{eqnarray}
|S_{\mathcal{L}} - S_{\mathcal{E}\mathcal{F}_{\mathcal{E}}}| \leq S_{\mathcal{L}\mathcal{E}\mathcal{F}_{\mathcal{E}}} \leq S_{\mathcal{L}} + S_{\mathcal{E}\mathcal{F}_{\mathcal{E}}},
\end{eqnarray}
we obtain that $S_{\mathcal{L}\mathcal{E}\mathcal{F}_{\mathcal{E}}} = S_{\mathcal{L}}$.
\item[-- L3$'$.] By plugging L1$'$ and L2$'$ to the strong subadditivity relation $S_{\mathcal{L}\mathcal{E}} + S_{\mathcal{E}\mathcal{F}_{\mathcal{E}}} \geq S_{\mathcal{E}} + S_{\mathcal{L}\mathcal{E}\mathcal{F}_{\mathcal{E}}}$, we obtain the relation
\begin{eqnarray}
S_{\mathcal{E}} > S_{\mathcal{E}} + S_{\mathcal{L}},
\end{eqnarray}
where this is impossible.
\end{itemize}
Therefore, we can repeat the similar thought experiment of AMPS in the mirror case.

The physical meaning is clear. In the black hole case, before and after the Page time, due to the unitarity, there is a pair of maximally entangled particles, where one is in the later part of Hawking radiation (let us call it $\alpha$) and the other is in the earlier part of Hawking radiation (let us call it $\beta$). At the same time, according to the semi-classical quantum field theory, there is an entanglement between $\alpha$ and its counterpart $\alpha'$. However, this is impossible, since $\alpha$ is impossible to be entangled with both $\beta$ and $\alpha'$ at the same time.

For the mirror case, due to the trajectory of the mirror, there is no $\alpha'$. However, there is an accessible counterpart of $\beta$ that we can call it $\beta'$. Therefore, now $\beta$ causes an inconsistency since $\beta$ is entangled with both $\beta'$ and $\alpha$. This is a simple explanation of the inconsistency of the moving mirror.

\subsection{Then, what happens?}

If this inconsistency is real, then what is wrong? Since there is a tension between the unitary quantum mechanics (i.e., the entanglement entropy formula) and the semi-classical quantum field theory (i.e., the radiation formula according to the mirror trajectory), there are basically two possibilities. The first possibility is that our entanglement entropy formula is wrong and even the unitarity itself would not be preserved. The other possibility is that the radiation formula breaks down after the Page time.

The latter possibility may be a more conservative assumption. The Davies-Fulling-Unruh formula \cite{Davies:1976ei} is true in the context of the semi-classical analysis. However, quantum mechanically, there should be a partner mode of the out-going flux \cite{Hotta:2015yla}, where the quantum interactions between the partner mode and the mirror itself are less clear in the semi-classical analysis. The interaction between the mirror and the partner mode will be turned on after the mirror begins to decelerate (i.e., after the Page time). If the interaction between the mirror and the in-going partner mode cannot be covered by the semi-classical analysis, then it is not so surprising that the interaction between the in-going partner mode and the mirror generates a violent effect.

At the theoretical level, this corresponds to a quantum gravitational effect, the so-called \textit{firewall phenomena} \cite{Almheiri:2012rt} that would be seen by an outside observer \cite{Hwang:2012nn}. In the experimental connection, it is impossible for tabletop experiments to reach the quantum gravitational scale, but such violent effects may indeed burn or break the moving mirror. Therefore, in the tabletop experiment, if our interpretations are correct, then we conservatively expect that \textit{there will be a firewall-like effect that burns the moving mirror}.

\section{\label{sec:dis}Discussion: what can we learn from mirror experiments?}

In this paper, we investigated the entropy and the information transmission process in two-dimensional moving mirrors. This can in principle be fabricated by future laser-plasma experiments \cite{Chen:2015bcg}. The task of this paper is to show a connection between the mirror experiments and the information loss problem of black holes.

By designing the mirror trajectory, we can mimic possible candidates of resolutions to the information loss problem. We can summarize as follows.
\begin{itemize}
\item[1.] A suddenly stopping mirror mimics the case that the final burst carries all information. However, this mirror analogy shows that in order to recover all information, the final burst should have a large amount of energy. Hence, we conclude that the final burst is not enough to carry all information in the realistic black hole evaporation.
\item[2.] A slowly stopping mirror mimics the case that thermal Hawking radiation carries information.
\item[3.] A long propagating mirror mimics a Planck scale remnant. Even though the remnant has a small amount of energy, as long as the lifetime is long enough, it can restore information. Therefore, this partly \textit{demonstrates} the possibility of the remnant scenario.
\end{itemize}

On the other hand, we also checked the apparent inconsistency between the scenarios based on the unitary quantum mechanics (the entanglement entropy formula) and the semi-classical quantum field theory (the Davies-Fulling-Unruh formula). This opens a possibility that the interaction between the in-going partner mode and the decelerating mirror can cause a violent effect that can be interpreted as the firewall-like phenomena. This can be definitely demonstrated by Chen-Mourou plasma-laser experiments \cite{Chen:2015bcg}. This should happen for all possible decelerating trajectories, and hence, the analysis of the previous paragraph would remain tentative.

For all cases, there is a period of \textit{rapid energy decrease} around the Page time, though it is unclear whether we can detect the negative energy flux or not \cite{Good:2015nja,Abdolrahimi:2015tha}. In any case, such a rapid energy drop can be a clear expectation from theoretical calculations that needs to be confirmed by future experiments.

One additional question for the future investigation is as follows: \textit{how can we measure the entanglement entropy in realistic experiments}? This should be in principle possible, but it is fair to say that it will be a challenge. If we can find a way to measure the entanglement entropy, then we can confirm or falsify the renormalized entanglement entropy formula experimentally. Of course, even if the formula is not very correct, we reasonably guess that the qualitative properties of the entanglement entropy should not be so different from the formula, and hence experimental results should not be very different from the present paper. We leave these topics as future tasks that need to be confirmed by upcoming experiments.

\section*{Acknowledgment}
The authors would like to thank Bill Unruh for critical comments about this work. We also thank to Michael Good and Yen Chin Ong for valuable suggestions. DY is supported by Leung Center for Cosmology and Particle Astrophysics (LeCosPA) of National Taiwan University (103R4000).

\end{document}